\documentclass[preprint]{vgtc}

\onlineid{1163}

\vgtccategory{Research}

\preprinttext{Accepted for presentation at IEEE Visualization (VIS) 2024 short paper track and will appear in the conference proceedings.}

\title{Curve Segment Neighborhood-based Vector Field Exploration}

\author{Nguyen K.\ Phan\thanks{e-mail:nguyenpkk95@gmail.com} %
\and Guoning Chen\thanks{e-mail:gchen16@uh.edu}}
\affiliation{\scriptsize University of Houston}

\abstract{%

Integral curves have been widely used to represent and analyze various vector fields. 
In this paper, we propose a Curve Segment Neighborhood Graph (CSNG) to capture the relationships between neighboring curve segments. This graph representation enables us to adapt the fast community detection algorithm, i.e., the Louvain algorithm, to identify individual graph communities from CSNG. Our results show that these communities often correspond to the features of the flow. To achieve a multi-level interactive exploration of the detected communities, we adapt a force-directed layout that allows users to refine and re-group communities based on their domain knowledge. We incorporate the proposed techniques into an interactive system to enable effective analysis and interpretation of complex patterns in large-scale integral curve datasets.

}

\keywords{Vector field, neighbor search, community detection}

\graphicspath{{figs/}{figures/}{pictures/}{images/}{./}} 

\usepackage{tabu}                      
\usepackage{booktabs}                  
\usepackage{lipsum}                    
\usepackage{mwe}                       

\usepackage{mathptmx}                  
\usepackage{float}

\begin{document}


\setlength{\baselineskip}{0.975 \baselineskip}
\setlength{\abovedisplayskip}{1pt}
\setlength{\belowdisplayskip}{1pt}
\setlength{\abovedisplayshortskip}{3pt}
\setlength{\belowdisplayshortskip}{-2pt}
\setlength{\belowcaptionskip}{3pt}
\setlength{\abovecaptionskip}{2pt}
\setlength{\textfloatsep}{9pt}
\setlength{\floatsep}{3pt}
\setlength{\intextsep}{2pt}

\maketitle

\section{Introduction}
\label{sec:intro}

Integral curves are often applied to visualize and interpret vector field data. Depending on the seeding and placement strategy adopted and the complexity of the original data, integral curves may occlude each other and have varying densities in different regions. 
The large number of curves and segments further complicate their analysis.
Usually, to highlight meaningful features from sets of integral curves to aid their analysis, clustering-based techniques \cite{YuHierarchicalClustering,BloodFlow2014,nguyen2021physics,shi2021integral} or the pattern search approaches \cite{lu2013exploring,FlowString2016,Wang2016Patterns} can be applied. However, most curve-based clustering methods classify the entire curve, while at the same time, not all parts of a curve belong to a feature. Clustering-based methods also require specifying proper similarity metrics to produce meaningful results \cite{shi2021integral}. In the meantime, pattern search methods require the user to specify a reference pattern (or template) along with a similarity threshold for searching, which is highly dependent on the user's knowledge and the input data. Also, other interesting patterns than the reference pattern may not be highlighted. More importantly, pattern search need not support a level-of-detailed exploration of the patterns in the input integral curves.

\noindent \textbf{Our contribution. }To address the above challenges, we propose to represent the neighboring relations among segments of the input integral curves as a graph, called the \emph{Curve Segment Neighborhood Graph or (CSNG)}. The graph nodes correspond to the individual segments decomposed from the input curves, while the graph edges indicate the corresponding segments are neighboring to each other. 
With CSNG, we can adapt the community detection algorithms from graph analysis to group similar segments to form clusters for the first time. Our results show that the communities found in CSNG often correspond to meaningful patterns and features in the data.
In addition, by taking advantage of the fast computation of the Louvain community detection algorithm \cite{blondel2008fast,lancichinetti2009community}, we achieve real-time community detection, setting the foundation for the subsequent level-of-detail exploration of the integral curves.

To support an interactive exploration of the integral curves, we adapt the force-directed layout to visualize the detected communities from CSNG. Our force-directed layout allows the user to refine each community and re-group certain sub-community with another community or sub-community to form a new community. This addresses the need to interactively edit and correct the misclassification of segments. In addition, we develop a web-based interactive CSNG-centric exploration system for curve-based data. 
It incorporates the interactive, multi-layered force-directed graph enabled by graph community detection to provide a powerful tool for curve data analysis.

We have applied the proposed methods and the developed interactive visualization system to several streamline data sets. 
Our experiments show that our method and system allow the user to effectively explore complex 3D patterns. The users can also incorporate their domain knowledge to refine and modify the results interactively to achieve a detailed analysis of those patterns.

\section{Related Work}
\label{sec:related_work}

\noindent \textbf{Integral curve seeding, placement, and visualization. }
Depending on different goals, the seeding and placement strategy can lead to evenly-spaced streamlines in either the physical (object) space \cite{jobard1997creating,liu2006advanced,mebarki2005farthest,chen2007similarity}  or the image space \cite{turk1996image,spencer2009evenly,li2007image}, feature enhanced streamlines \cite{verma2000flow,chen07,wu2009topology}, and streamlines that best depict the flow based on the information theory \cite{FlowInfo10}. A recent survey paper \cite{sane2020survey} provides a detailed review of the existing integral curve seeding and placement strategies. 
Various visualization \cite{gunther2013opacity,tong2015view}, view-point selection \cite{LeeMSC11,tao2012unified}, and novel data structure representation \cite{lu2021curve} for integral curves have been proposed to support the exploration and interpretation of flow behaviors.

\noindent \textbf{Integral curve clustering. }
Clustering techniques \cite{shi2021integral} are often applied to group similar integral curves based on certain similarity criteria to highlight meaningful patterns to aid their interpretation. 
Various methods have been introduced, including hierarchical bundles of streamlines \cite{YuHierarchicalClustering}, pathline clustering \cite{nguyen2021physics}, machine learning based clustering \cite{han2018flownet}. Clustering method has also been used to aid the analysis of blood flow \cite{BloodFlow2014}, identify hairpin vortices \cite{zafar2023extract}, and fiber tracks in diffusion tensor imaging \cite{moberts2005evaluation,everts2015exploration}.

\noindent \textbf{Pattern search in integral curves.}
Different from clustering techniques, pattern search aims to identify curves or segments that possess similar characteristics to a user-specified reference.
Wang et al. \cite{wang2014pattern, wang2015multi, Wang2016Patterns} developed methodologies to identify patterns within vector field data. 
Lu et al. \cite{lu2013exploring} introduced a distribution based approach to characterize the streamline characteristics to facilitate their search and analysis.
Tao et al. \cite{Flowstring2014,FlowString2016} proposed to encode the streamline characteristics into character strings to facilitate pattern search.

While clustering and pattern search methods help analyze curve-based data, they either do not scale well to large-scale data or rely on specified references for analysis.

\noindent \textbf{Graph-based flow visualization and exploration. }
Xu et al. introduced Flow Web \cite{xu2010flow}, a graph-based user interface where nodes represent regions in the field and links connect regions with particle travel between them. This approach allows for systematic exploration of 3D flow data by minimizing occlusion and facilitating queries on flow properties. Xu and Shen proposed FlowGraph \cite{ma2013flowgraph}, a compound hierarchical graph representation that organizes streamline clusters and spatial regions hierarchically.

\begin{figure}[!t]
\centering
  \includegraphics[width=1\linewidth]{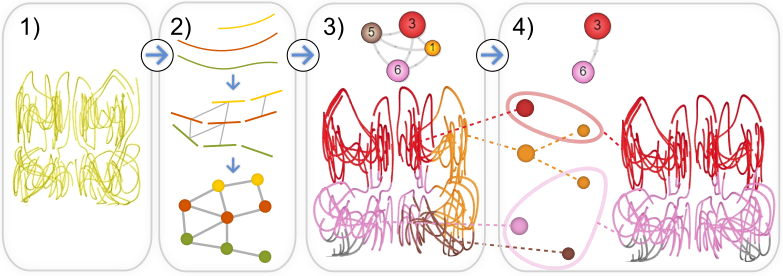}
   \caption{Illustration of our framework. 1) Load the streamline dataset. 2) Perform curve decomposition, then CSNG construction with either KNN or RBN. 3) Perform community detection on the CSNG to categorize each curve segment into community clusters. 4) Manually adjust the community results such as merging and splitting communities to fix misclassification.
   }
  \label{fig:pipeline}
\end{figure}

\section{Our Method}

Our method consists of a few important steps as illustrated in \autoref{fig:pipeline}. First, it represents the input curves and their relations by a directed graph. Second, it performs community detection on the graph to identify individual segment clusters. Third, the user explores and interprets patterns in the curve data in a multi-level fashion through a force-directed layout representation of the obtained communities.  
In the following sections, we provide a detailed description of how these steps are achieved. 

\subsection{Curve-centered Neighbor Search}
\label{sec:curveneighborsearch}

Our goal is to find the neighboring curve segments of a specific segment. We refer to this process as the \emph{curve-centered neighbor search}. To achieve that, we first partition each 3D (integral) curve into a set of curve segments. There are different strategies to decompose a 3D curve into smaller segments. Most of these strategies \cite{shi2021integral, lu2021curve} rely on curvature information to determine cut points. Since most integral curve data sets we experiment with were generated using constant integration step sizes, we opt for a simple strategy to achieve a multi-resolution neighbor search. Intuitively, we consider every $L$ integral line segment along a 3D curve to form a curve segment. $L$ is therefore the adjustable resolution of our neighbor search framework. The highest resolution is $L=1$ where each streamline segment is considered a curve segment for the neighbor search. Lower resolution (i.e., with a larger $L$ value) will speed up the neighborhood finding process but may significantly decrease the information density of the resulting set of neighbors, resulting in errors in the neighbor search.

\noindent \textbf{Distance metrics. }
Consider a query segment $C$ and a candidate segment $l_i$. To determine their neighboring relation, their distance, $\mathbf{d}(L_i,C)$, needs to be computed.
Three distance metrics can be considered, i.e., the shortest, the longest, and the average distance. In this work, \emph{we select the longest distance}. This is because our simple decomposition leads to straight-line segments, the longest distance between them is equivalent to their Hausdorff distance. Also, the shortest distance may select segments $l_i$ that only have one endpoint close to $C$ while the rest is pointing away from $C$. In this case, most parts of the selected segments are not close to $C$. The average distance partially addresses the limitation of the shortest distance, but it is more expensive to compute.

\noindent \textbf{Neighbor search strategies. }
We consider two neighbor search strategies, i.e., K-nearest neighbor search and radius-based neighbor search. Based on a selected distance metric $\mathbf{d}(L_i,C)$, the K-nearest neighbor search (KNN) ranks the segments in ascending order based on their distance to $C$ and selects the top $K$ segments as the neighbors. Similarly, the radius-based neighbor search (RBN) identified all segments whose distances to $C$ are smaller than a threshold $R$ as the neighbors of $C$. 
In our implementation, we use a standard segment-based KD-tree \cite{arya1998optimal} to organize the individual segments to efficiently identify candidate segments for our neighbor search.

\noindent \textbf{Construct CSNG. }
To represent the neighboring relations among segments, we construct the \textbf{Curve Segment Neighborhood Graph (CSNG)}. 
A CSNG is a \emph{directed graph} \( G_{CS} = (V_{CS}, E_{CS}) \), which comprises graph nodes \( V_{CS} \) representing individual curve segments, and directed edges \( E_{CS} \) indicating neighbor relationships. 
Properties of the segments (e.g., curvature, length, and velocity magnitude) and the difference of these properties between neighboring segments can be stored on the nodes and edges of CSNG as weights. These attributes inform the strength and similarity of connections.
Note that a CSNG constructed based on KNN is a directed graph, while it can be an undirected graph with RBN.

Next, we extract meaningful patterns and features from the input 3D (integral) curves with the aid of CSNG. 
Traditional clustering and segmentation techniques, such as DBSCAN, K-Means, and agglomerative hierarchical clustering (AHC), often rely on pairwise distance calculations between segments, which can be computationally expensive for large datasets. Moreover, they struggle to incorporate the relational information between segments, which can be useful for understanding the underlying dynamics of the vector field. To address this, we borrow the community detection techniques to analyze CSNGs.

\subsection{Community Detection on CSNGs}
\label{sec:communitydetect}

Community detection aims to identify groups of nodes in a graph that are more densely connected internally than with the rest of the network. For CSNG, we define a community as a group of segments forming a cohesive cluster, representing features of the curve dataset

We employ the Louvain algorithm \cite{blondel2008fast,lancichinetti2009community} for community detection due to its efficiency and ability to uncover hierarchical community structures. 
The algorithm's resolution parameter controls the granularity of detected communities. A smaller resolution leads to coarser partitioning, while a larger value results in finer communities. However, the same resolution value may yield different numbers of communities across datasets due to variations in graph connectivity.
To leverage the Louvain algorithm effectively, we encode relevant properties into CSNG edge weights, such as the distance between line segments and their orientation difference.

\begin{figure}[!t]
\centering
\includegraphics[width=0.88\columnwidth]{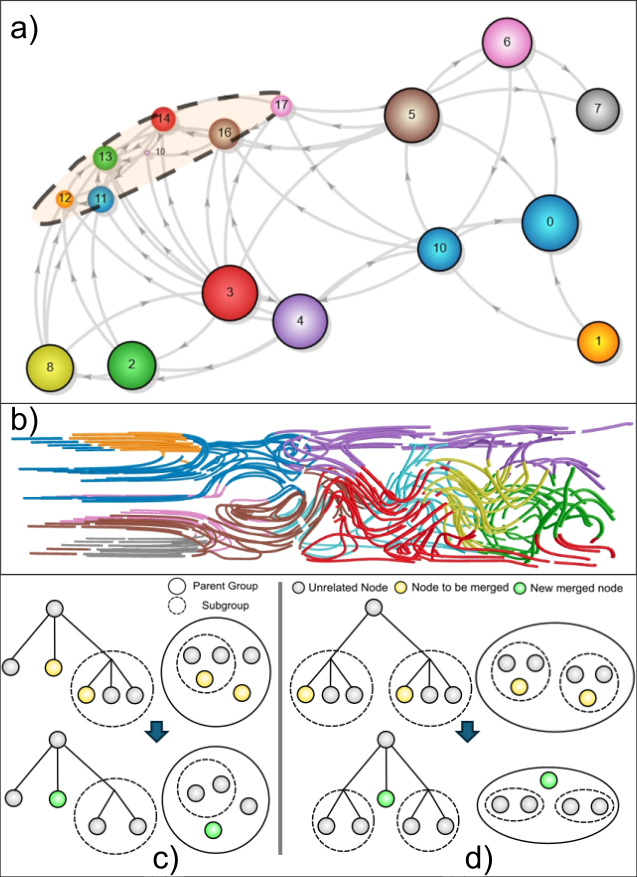}
\caption{
    a) The multi-layered force-directed graph layout, generated by applying the Louvain community detection algorithm (with a resolution of 1) to the Cylinder dataset. The group of nodes enclosed by a dashed outline highlights a split operation performed on a node using the Louvain method at a finer resolution of 0.3. b) Displays the same nodes represented as segment clusters in a 3D view. c) merging a sub-group with a parent group (yellow) results in the sub-groups being part of the parent group (green). (d) merging two sub-groups from two branches results in a new parent group.
}
\label{fig:forcegraph}
\end{figure}

\subsection{Visualize and Interact with CSNG Communities}
\label{sec:force-directed-graph}

We use a multi-layered force-directed graph to visualize CSNG and its communities in a hierarchical fashion.
Our force-directed graph is formulated as a compound graph \(G_f = (V_f, E_f)\), where \(V_f\) represents the set of nodes, each corresponding to a distinct community (or cluster) of curve segments obtained from community detection algorithms. The set \(E_f\) consists of edges connecting these nodes, with each edge \((v_i, v_j) \in E_f\) indicating a neighborhood relationship between segments in clusters \(v_i\) and \(v_j\). To encode the hierarchical information, each node may contain a set of sub-nodes corresponding to the sub-communities, and each edge could be a collection of edges connecting the corresponding sub-communities. Our edge spring force in the multi-layered graphs is inspired by Lu et al.'s work \cite{liu2022improved} and our node spring force in the clustered graphs adapts the work by Eades et al. \cite{eades2004navigating}. A detailed algorithm is provided in the supplemental document. We visualize a node with a sphere whose size is determined by the number of segments within the corresponding community. It receives a distinct color based on its community ID. We use arcs to depict the edges.

\noindent \textbf{Split or merge nodes in hierarchical communities. }
Nodes of our graph can be dynamically merged based on user interaction or predefined criteria, with the stipulation that for nodes from different groups to be merged, they must belong to the same hierarchical branch. This merging criterion is vital for preserving the inherent hierarchical structure of the graph. When it comes to maintaining hierarchical integrity, if the selected nodes \(V_f\) are from different groups, a merge is only allowed if these groups are on a common hierarchical path, ensuring the logical consistency of group relationships and the overall integrity of the graph's hierarchy. Furthermore, in terms of resulting group membership, the newly formed node post-merger inherits its group membership from the most encompassing group among the original vertices. Consequently, if a node from a subgroup is merged with a node from its parent group, the new node will belong to the parent group (Figure \ref{fig:forcegraph}(c)). In another case, if a node from a subgroup is merged with a node from a different subgroup, the new node will form a new node in the parent's group (Figure \ref{fig:forcegraph}(d)).
A detailed algorithm of how this is achieved can be found in the supplemental document.

\section{Results and Evaluation}
\label{sec:results}

We integrate the CSNG construction, community detection from CSNG, and the interactive exploration and modification of the community detection results into a web-based system  \footnote{https://github.com/MangoLion/CSN\_VIS}. Details of this system can be found in the supplemental document. 
We apply our method and the system to four streamline data sets computed from the Bernard convection, a flow behind a square cylinder, Plume simulation, and Crayfish simulation, respectively. 
We used a uniform seeding strategy for the Bernard convection, flow behind a square cylinder, and Plume simulation datasets, placing seed points in a grid-based uniform spacing across the 3D vector field. For the Crayfish simulation dataset, which contains large regions of low velocity, we employed random seeding to avoid over-sampling and dense bundling of streamlines. Streamline integration was performed using an RK4 integrator with a fixed step size, normalizing each segment vector for consistent length.


\begin{table}[b]
\centering
\caption{
Performance of Our Framework. For CSNG construction, the parameters are K=60 for KNN, R=10\% dataset bounds diagonal for RBN. Louvain resolution is 1. Due to size constraint the Plume dataset's segments were merged together by a ratio of 4:1.
}
\label{tab:performance}
\small
\begin{tabular}{|l|r|r|c|c|c|}
\hline
\textbf{Dataset} & \textbf{\# Lines} & \textbf{\# Segments} & \multicolumn{2}{c|}{\textbf{CSNG Duration (s)}} & \textbf{Louvain} \\
\cline{4-5}
 &  &  & \textbf{KNN} & \textbf{RBN} &  \textbf{Detection(s)}\\
\hline
Bernard & 128 & 12146 & 5.32 & 13.46 & 18.53 \\
Crayfish & 216 & 28913 & 8.66 & 24.02 & 7.4 \\
Plume* & 128 & 16077 & 6.32 & 21.12 & 3.32 \\
Cylinder & 250 & 7559 & 4.56 & 6.32 & 11.13 \\
\hline
\end{tabular}
\label{tab:performance}
\end{table}



\noindent \textbf{Performance.}
Table \ref{tab:performance} presents the execution times associated with transforming integral curves into CSNG directed graph data across four distinct datasets. 
The current processing is carried out on a system powered by a Ryzen 5 3600 CPU and 32GB of DDR4 2666 MHz RAM, operating under CPU-bound conditions. 

\begin{figure}[t]
\centering
  \includegraphics[width=0.92\linewidth]{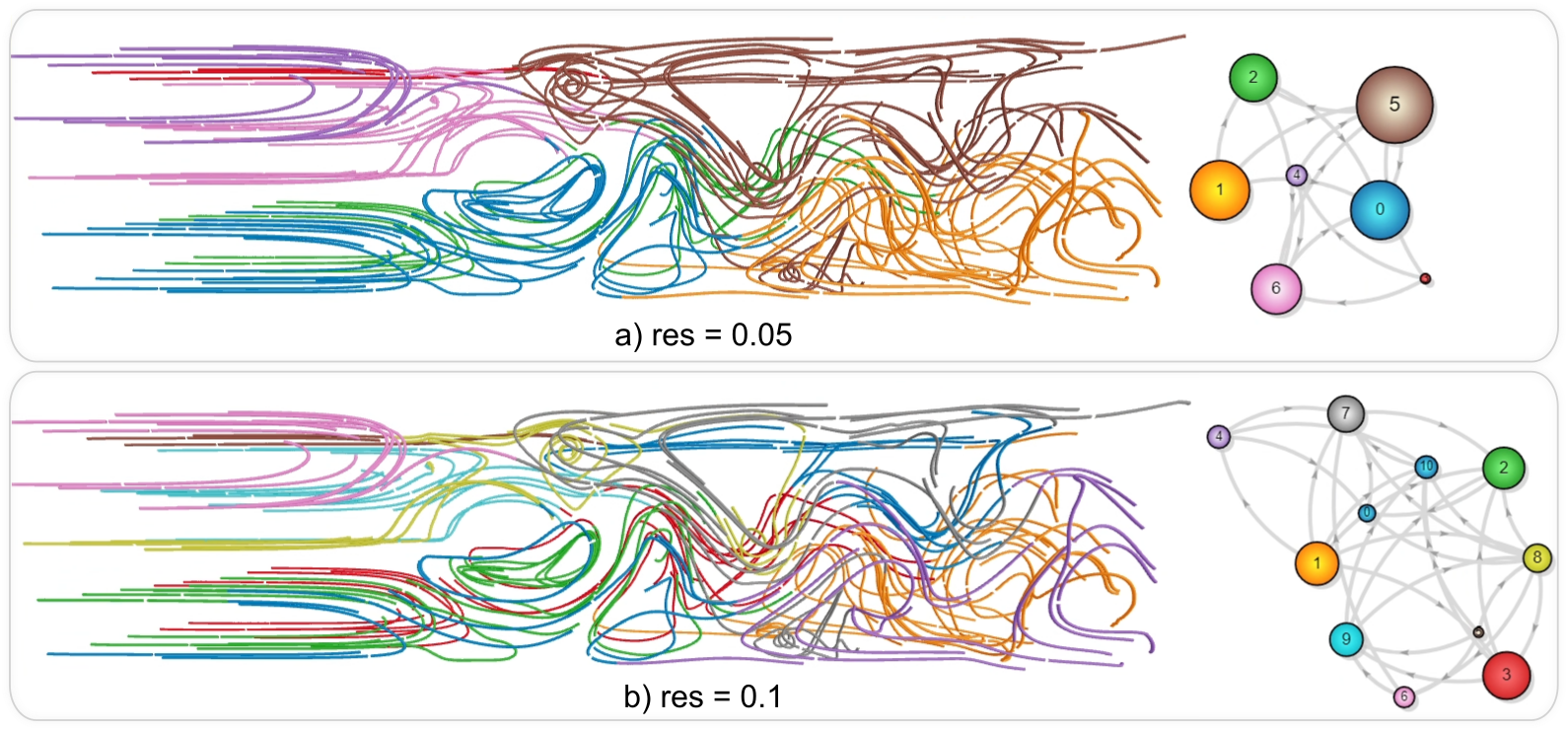}
   \caption{Impact of different values of the resolution parameter for community detection. (a) the community detection result on the cylinder data with a resolution value of 0.05 and (b) the result with a resolution value of 0.1. A larger resolution leads to a finer result.}
  \label{fig:resolution}
\end{figure}

\noindent \textbf{Impact of the resolution parameter.}
The resolution parameter of the Louvain algorithm can be used to indirectly control the granularity of the detected community. In general, a small resolution will lead to coarse community detection, and a large resolution will lead to a fine detection result. \autoref{fig:resolution} demonstrates the effect of the resolution on the community detection results.
However, it is worth noting that the same resolution value does not always lead to the same number of detected communities for different data sets. This is because the detected communities of a graph highly depend on the connectivity density of the graph. A sparse graph will result in more communities even if a small resolution value is used. Therefore, in practice, it is up to the user to tune the resolution value to achieve the desired granularity of the results. From our experiments, we found that a value less than 0.5 is a good start for the testing data.

\begin{figure}[!t]
\centering
  \includegraphics[width=1\linewidth]{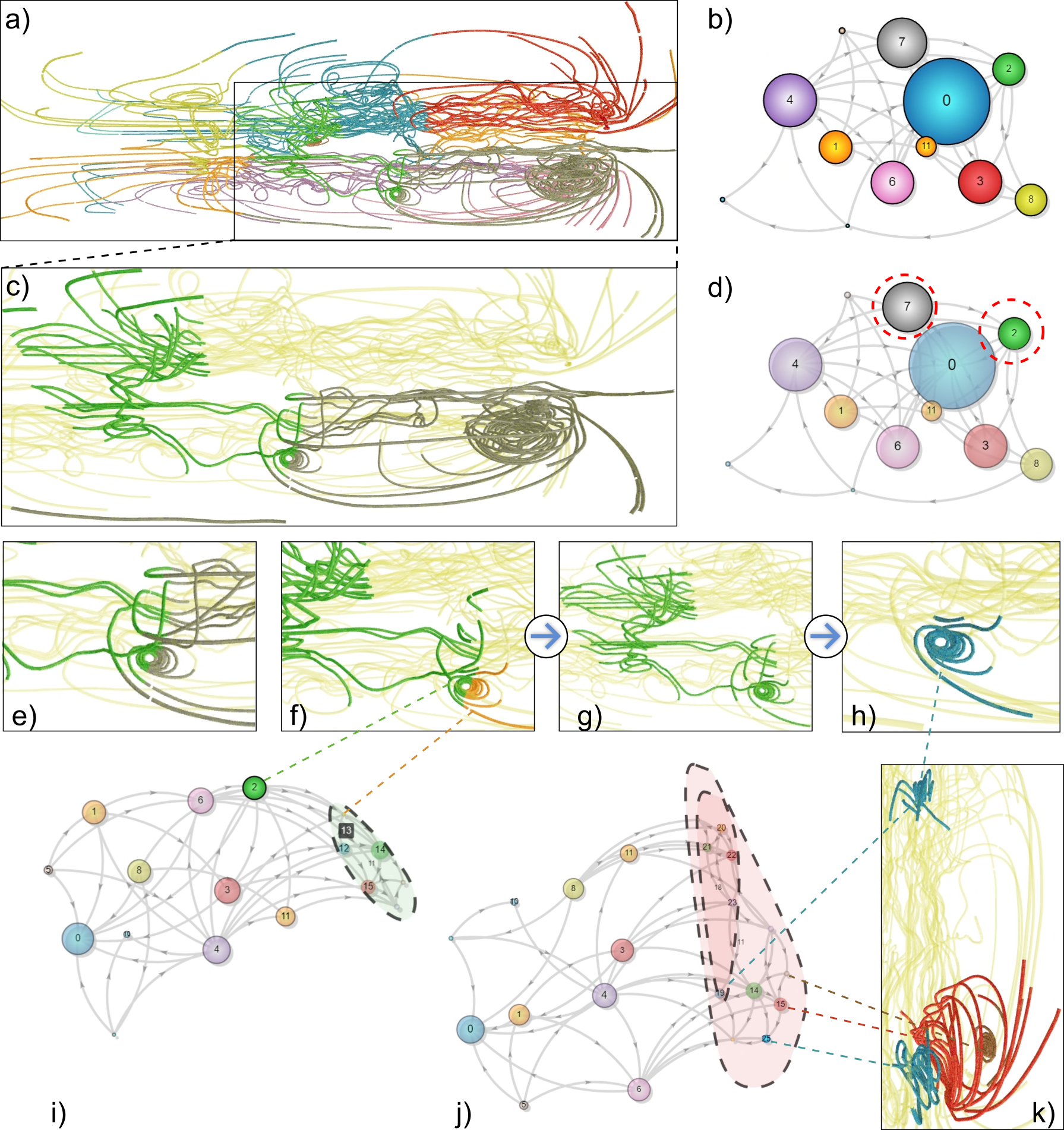}
   \caption{A level-of-detail analysis of the Plume streamline data set.}
  \label{fig:usecase}
\end{figure}

\noindent \textbf{A use case.}
To demonstrate how to use our method to explore and interpret curve-based data sets, we apply it to the Plume streamline data set (\autoref{fig:usecase}). 
The solar plume simulation models the behavior of a solar eruption, a large and bright feature extending outward from the Sun's surface.
 This data set contains 128 streamlines (\autoref{fig:usecase}(a)). After decomposition with $L=2$, we obtain 32191 segments. 
Next, we select KNN with K=60 to construct a CSNG. 
We apply the Louvain algorithm with resolution value = 0.5 to detect the initial communities from the obtained CSNG, which yields 15 communities (\autoref{fig:usecase}(b)). 
We focus on communities 7 and 2, which are highlighted in both the 3D volume rendering and the force-directed graph views (\autoref{fig:usecase}(c) and (d)). These two communities were specifically selected because they contain all of the visually interesting vortex-like features that spanned a significant portion, approximately half, of the frontal part of the plume data set. 
A detailed inspection of these two communities reveals a vortex feature that is misclassified as belonging to both groups, as indicated in \autoref{fig:usecase}(e). The misclassified region appears as an overlapping region between communities 7 and 2 in the force-directed graph view, suggesting that the vortex feature is incorrectly split between the two communities. To resolve this misclassification, we employ the following steps:
(1) We closely inspect the misclassified area and select all communities involved in the classification. In this case, communities 7 and 2 are selected, as shown in \autoref{fig:usecase}(e). (2) We perform a subgroup splitting of community 7 using the Louvain algorithm with a resolution parameter of 0.5. This step divides the large community into 6 smaller subgroups (\autoref{fig:usecase}(i)). (3) We identify the subgroup that contains the misclassified vortex feature (\autoref{fig:usecase}(i) subgroup 13) and merge it into community 2 to create a more coherent representation of the feature (\autoref{fig:usecase}(h)). This step ensures that the vortex feature is no longer split across multiple communities. (4) Finally, we perform another split on the merged community 2 using the Louvain algorithm with a resolution of 0.5. This step breaks apart the merged community, revealing the vortex feature in its entirety as its distinct subgroup (\autoref{fig:usecase}(h) and (j)). The force-directed graph view (\autoref{fig:usecase}(j)) and the 3D segment view (\autoref{fig:usecase}(k)) showcase the final result of the community refinement process. The vortex feature is now clearly separated from the other subgroups within communities 7 and 2, which have been split into a total of 13 smaller subgroups. In addition, all relevant features, represented as four subgroups, are each highlighted on both views, illustrating the distinct vortex features within the original communities 7 and 2. Two convex envelopes with a dashed boundary are used in the force-directed graph view to encapsulate these subgroups, further highlighting the refined community structure. This use case demonstrates the effectiveness of our approach in resolving misclassifications and extracting meaningful features from complex data sets like the Plume.



\begin{figure}[!t]
\centering
  \includegraphics[width=0.9\linewidth]{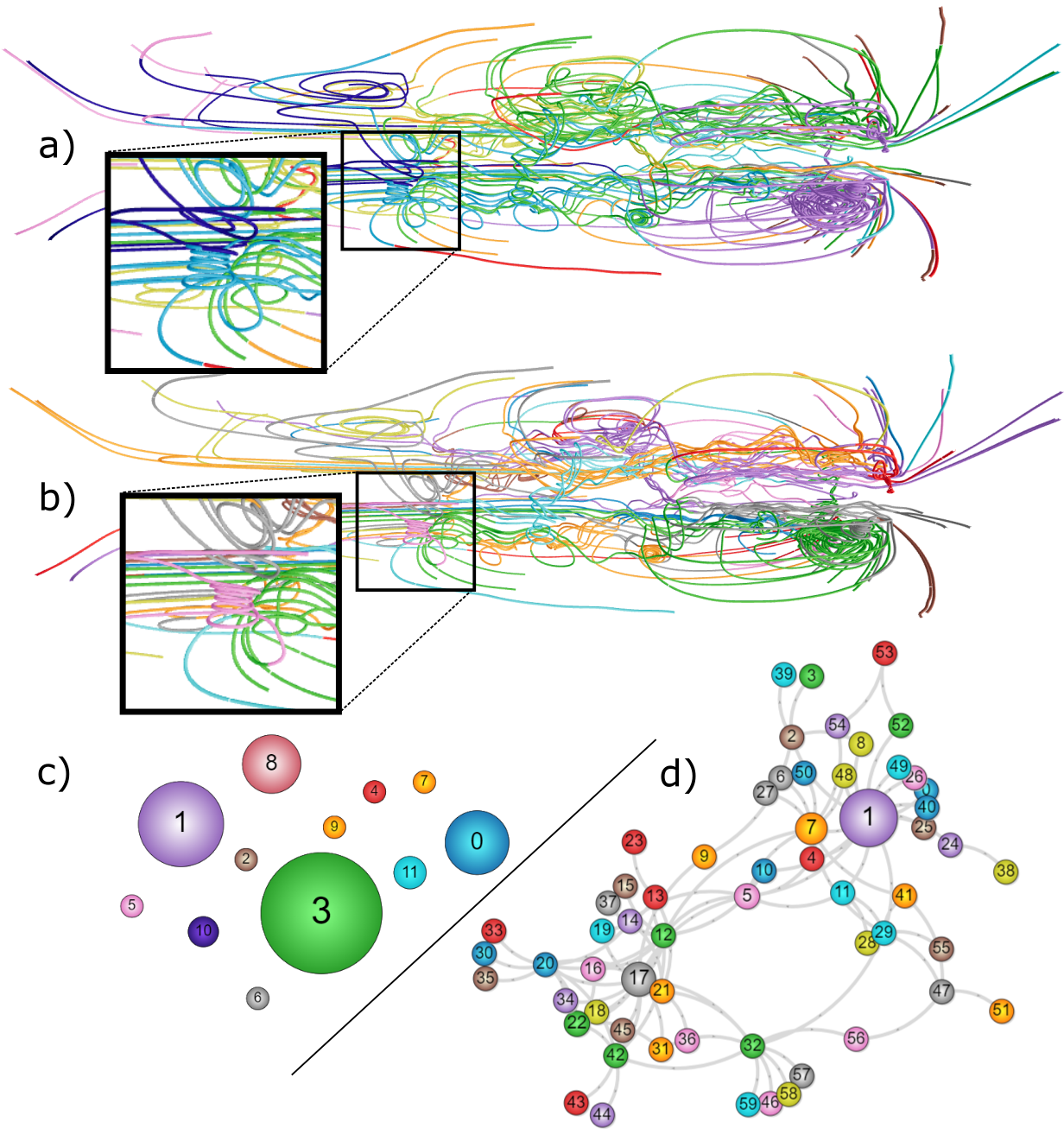}
   \caption{Detailed comparison of the best clustering results for the Plume dataset using PCA K-means and Louvain algorithms. PCA K-means parameters: dim=5, k=12; Louvain parameters: resolution=1, RBN radius=1\%. 3D rendering of PCA K-means clustering result and community detection result are shown in (a) and (b), respectively. (c) and (d) are force directed graphs of (a) and (b).}
  \label{fig:pcacompare}
\end{figure}

\noindent \textbf{Comparison with PCA k-means clustering.}

We compare our method with the PCA k-means clustering, which has been shown to perform well for integral curve clustering \cite{shi2021integral}. Figure \ref{fig:pcacompare} shows such a comparison on the Plume dataset.
{For PCA k-means, we used 5 principal components and k=12 clusters, which produced the best visual results after parameter tuning. For Louvain community detection, we used a resolution of 1 and RBN with a radius of 1\% of the dataset bounds.  
As seen in Figure \ref{fig:pcacompare}a, PCA k-means identifies vortex-like structures but often splits them across multiple clusters, as evidenced by the multiple colors in the zoomed inset. In contrast, our Louvain-based approach (Figure \ref{fig:pcacompare}b) yields more consistent community detection results, identifying coherent structures as belonging to single communities, as shown by the more uniform coloring in the inset.
Figures \ref{fig:pcacompare}c and \ref{fig:pcacompare}d provide force-directed graph layouts of the clustering results, further illustrating the differences between the two approaches. The PCA k-means result (Figure \ref{fig:pcacompare}c) shows 12 distinct clusters represented by differently colored and sized nodes. The Louvain result (Figure \ref{fig:pcacompare}d) reveals a more complex, connective community structure with nodes of various sizes representing communities and sub-communities, and edges showing relationships between them.
This comparison shows that while PCA k-means can still identify meaningful clusters, our approach offers advantages in terms of feature coherence and the ability to interactively refine communities hierarchically. This flexibility is particularly valuable for exploring complex flow structures where the optimal segmentation may not be immediately apparent. In terms of runtime, PCA k-means (15.67s) outperformed our RBN Louvain approach (29.12s), mainly due to the overhead of RBN neighbor search. However, the clustering phases alone (PCA k-means vs. Louvain) showed similar times (15.67s vs. 16.2s).




\section{Conclusion and Future Work}
\label{sec:conclusion}

In this work, we proposed to use a direct graph, called CSNG, to represent the neighboring relation among curve segments of the input curve-based data set. 
This allows us to adapt the fast community detection algorithm for analyzing CSGNs, achieving the clustering of the segments. We also developed a multi-layered force-directed layout technique for the detected communities to support a multi-level exploration of the patterns in the input curves. We implemented our method in a web-based system to support an interactive exploration of various curve-based data. We have applied our method and system to a few integral curve data to evaluate its effectiveness. 

While our current implementation shows promise, we identified limitations in scaling to very large numbers of segments due to memory inefficiency in our web worker implementation. It also needs to study the impact of RBN and KNN on the constructed CSNG and the subsequent analysis, and a thorough evaluation of the proposed method. We plan to address these in the future work.

\acknowledgments{
This research was supported by NSF OAC 2102761.}

\bibliographystyle{abbrv-doi-hyperref}

\bibliography{Content/references}

\begin{thebibliography}{10}

\bibitem{arya1998optimal}
S.~Arya, D.~M. Mount, N.~S. Netanyahu, R.~Silverman, and A.~Y. Wu.
\newblock An optimal algorithm for approximate nearest neighbor searching fixed dimensions.
\newblock {\em Journal of the ACM (JACM)}, 45(6):891--923, 1998.

\bibitem{blondel2008fast}
V.~D. Blondel, J.-L. Guillaume, R.~Lambiotte, and E.~Lefebvre.
\newblock Fast unfolding of communities in large networks.
\newblock {\em Journal of statistical mechanics: theory and experiment}, 2008(10):P10008, 2008.

\bibitem{chen07}
G.~Chen, K.~Mischaikow, R.~S. Laramee, P.~Pilarczyk, and E.~Zhang.
\newblock {Vector Field Editing and Periodic Orbit Extraction Using Morse Decomposition}.
\newblock {\em IEEE Transactions on Visualization and Computer Graphics}, 13(4):769--785, Jul./Aug. 2007.

\bibitem{chen2007similarity}
Y.~Chen, J.~Cohen, and J.~Krolik.
\newblock Similarity-guided streamline placement with error evaluation.
\newblock {\em IEEE Transactions on Visualization and Computer Graphics}, 13(6):1448--1455, 2007.

\bibitem{eades2004navigating}
P.~Eades and M.~L. Huang.
\newblock Navigating clustered graphs using force-directed methods.
\newblock In {\em Graph Algorithms And Applications 2}, pp. 191--215. World Scientific, 2004.

\bibitem{everts2015exploration}
M.~H. Everts, E.~Begue, H.~Bekker, J.~B. Roerdink, and T.~Isenberg.
\newblock Exploration of the brain’s white matter structure through visual abstraction and multi-scale local fiber tract contraction.
\newblock {\em IEEE transactions on visualization and computer graphics}, 21(7):808--821, 2015.

\bibitem{gunther2013opacity}
T.~G{\"u}nther, C.~R{\"o}ssl, and H.~Theisel.
\newblock Opacity optimization for 3d line fields.
\newblock {\em ACM Transactions on Graphics (TOG)}, 32(4):1--8, 2013.

\bibitem{han2018flownet}
J.~Han, J.~Tao, and C.~Wang.
\newblock Flownet: A deep learning framework for clustering and selection of streamlines and stream surfaces.
\newblock {\em IEEE transactions on visualization and computer graphics}, 26(4):1732--1744, 2018.

\bibitem{jobard1997creating}
B.~Jobard and W.~Lefer.
\newblock Creating evenly-spaced streamlines of arbitrary density.
\newblock In {\em Visualization in Scientific Computing’97}, pp. 43--55. Springer, 1997.

\bibitem{lancichinetti2009community}
A.~Lancichinetti and S.~Fortunato.
\newblock Community detection algorithms: a comparative analysis.
\newblock {\em Physical review E}, 80(5):056117, 2009.

\bibitem{LeeMSC11}
T.-Y. Lee, O.~Mishchenko, H.-W. Shen, and R.~Crawfis.
\newblock View point evaluation and streamline filtering for flow visualization.
\newblock In G.~D. Battista, J.-D. Fekete, and H.~Qu, eds., {\em PacificVis}, pp. 83--90. IEEE, 2011.

\bibitem{li2007image}
L.~Li and H.-W. Shen.
\newblock Image-based streamline generation and rendering.
\newblock {\em IEEE Transactions on Visualization and Computer Graphics}, 13(3):630--640, 2007.

\bibitem{liu2022improved}
J.~Liu, Z.~Wang, J.~Yu, H.~Liu, and T.~Li.
\newblock An improved force-directed automatic layout method for undirected compound graphs.
\newblock pp. 242--249, 2022.

\bibitem{liu2006advanced}
Z.~Liu, R.~Moorhead, and J.~Groner.
\newblock An advanced evenly-spaced streamline placement algorithm.
\newblock {\em IEEE Transactions on Visualization and Computer Graphics}, 12(5):965--972, 2006.

\bibitem{lu2013exploring}
K.~Lu, A.~Chaudhuri, T.~Y. Lee, H.~W. Shen, and P.~C. Wong.
\newblock Exploring vector fields with distribution-based streamline analysis.
\newblock In {\em 2013 IEEE Pacific Visualization Symposium (PacificVis)}, pp. 257--264, Feb 2013. \href{https://doi.org/10.1109/PacificVis.2013.6596153}
{doi: {{%
10\hspace{.1pt}\discretionary{.}{%
}{.}\hspace{.4pt}1109\discretionary{/}{%
}{/}PacificVis\hspace{.1pt}\discretionary{.}{%
}{.}\hspace{.4pt}2013\hspace{.1pt}\discretionary{.}{%
}{.}\hspace{.4pt}6596153}}}


\bibitem{lu2021curve}
Y.~Lu, L.~Cheng, T.~Isenberg, C.-W. Fu, G.~Chen, H.~Liu, O.~Deussen, and Y.~Wang.
\newblock Curve complexity heuristic kd-trees for neighborhood-based exploration of 3d curves.
\newblock In {\em Computer Graphics Forum}, vol.~40, pp. 461--474. Wiley Online Library, 2021.

\bibitem{ma2013flowgraph}
J.~Ma, C.~Wang, and C.-K. Shene.
\newblock Flowgraph: A compound hierarchical graph for flow field exploration.
\newblock In {\em 2013 IEEE Pacific Visualization Symposium (PacificVis)}, pp. 233--240. IEEE, 2013.

\bibitem{mebarki2005farthest}
A.~Mebarki, P.~Alliez, and O.~Devillers.
\newblock Farthest point seeding for efficient placement of streamlines.
\newblock In {\em VIS 05. IEEE Visualization, 2005.}, pp. 479--486. IEEE, 2005.

\bibitem{moberts2005evaluation}
B.~Moberts, A.~Vilanova, and J.~J. Van~Wijk.
\newblock Evaluation of fiber clustering methods for diffusion tensor imaging.
\newblock In {\em IEEE Visualization Conference}, pp. 65--72. IEEE Computer Society, 2005.

\bibitem{nguyen2021physics}
D.~B. Nguyen, L.~Zhang, R.~S. Laramee, D.~Thompson, R.~O. Monico, and G.~Chen.
\newblock Physics-based pathline clustering and exploration.
\newblock {\em Computer Graphics Forum}, 40(1):22--37, 2021.

\bibitem{BloodFlow2014}
S.~{Oeltze}, D.~J. {Lehmann}, A.~{Kuhn}, G.~{Janiga}, H.~{Theisel}, and B.~{Preim}.
\newblock Blood flow clustering and applications invirtual stenting of intracranial aneurysms.
\newblock {\em IEEE Transactions on Visualization and Computer Graphics}, 20(5):686--701, May 2014. \href{https://doi.org/10.1109/TVCG.2013.2297914}
{doi: {{%
10\hspace{.1pt}\discretionary{.}{%
}{.}\hspace{.4pt}1109\discretionary{/}{%
}{/}TVCG\hspace{.1pt}\discretionary{.}{%
}{.}\hspace{.4pt}2013\hspace{.1pt}\discretionary{.}{%
}{.}\hspace{.4pt}2297914}}}


\bibitem{sane2020survey}
S.~Sane, R.~Bujack, C.~Garth, and H.~Childs.
\newblock A survey of seed placement and streamline selection techniques.
\newblock In {\em Computer Graphics Forum}, vol.~39, pp. 785--809. Wiley Online Library, 2020.

\bibitem{shi2021integral}
L.~Shi, R.~Laramee, and G.~Chen.
\newblock Integral curve clustering and simplification for flow visualization: A comparative evaluation.
\newblock {\em IEEE transactions on visualization and computer graphics}, 27(3):1967 -- 1985, 2021.

\bibitem{spencer2009evenly}
B.~Spencer, R.~S. Laramee, G.~Chen, and E.~Zhang.
\newblock Evenly spaced streamlines for surfaces: An image-based approach.
\newblock In {\em Computer Graphics Forum}, vol.~28, pp. 1618--1631. Wiley Online Library, 2009.

\bibitem{tao2012unified}
J.~Tao, J.~Ma, C.~Wang, and C.-K. Shene.
\newblock A unified approach to streamline selection and viewpoint selection for 3d flow visualization.
\newblock {\em IEEE Transactions on Visualization and Computer Graphics}, 19(3):393--406, 2012.

\bibitem{Flowstring2014}
J.~Tao, C.~Wang, and C.~K. Shene.
\newblock Flowstring: Partial streamline matching using shape invariant similarity measure for exploratory flow visualization.
\newblock In {\em 2014 IEEE Pacific Visualization Symposium}, pp. 9--16, March 2014. \href{https://doi.org/10.1109/PacificVis.2014.12}
{doi: {{%
10\hspace{.1pt}\discretionary{.}{%
}{.}\hspace{.4pt}1109\discretionary{/}{%
}{/}PacificVis\hspace{.1pt}\discretionary{.}{%
}{.}\hspace{.4pt}2014\hspace{.1pt}\discretionary{.}{%
}{.}\hspace{.4pt}12}}}


\bibitem{FlowString2016}
J.~Tao, C.~Wang, C.~K. Shene, and R.~A. Shaw.
\newblock A vocabulary approach to partial streamline matching and exploratory flow visualization.
\newblock {\em IEEE Transactions on Visualization and Computer Graphics}, 22(5):1503--1516, May 2016. \href{https://doi.org/10.1109/TVCG.2015.2440252}
{doi: {{%
10\hspace{.1pt}\discretionary{.}{%
}{.}\hspace{.4pt}1109\discretionary{/}{%
}{/}TVCG\hspace{.1pt}\discretionary{.}{%
}{.}\hspace{.4pt}2015\hspace{.1pt}\discretionary{.}{%
}{.}\hspace{.4pt}2440252}}}


\bibitem{tong2015view}
X.~Tong, J.~Edwards, C.-M. Chen, H.-W. Shen, C.~R. Johnson, and P.~C. Wong.
\newblock View-dependent streamline deformation and exploration.
\newblock {\em IEEE transactions on visualization and computer graphics}, 22(7):1788--1801, 2015.

\bibitem{turk1996image}
G.~Turk and D.~Banks.
\newblock Image-guided streamline placement.
\newblock In {\em Proceedings of the 23rd annual conference on Computer graphics and interactive techniques}, pp. 453--460, 1996.

\bibitem{verma2000flow}
V.~Verma, D.~Kao, and A.~Pang.
\newblock A flow-guided streamline seeding strategy.
\newblock In {\em Proceedings Visualization 2000. VIS 2000 (Cat. No. 00CH37145)}, pp. 163--170. IEEE, 2000.

\bibitem{wang2014pattern}
Z.~Wang, J.~M. Esturo, H.-P. Seidel, and T.~Weinkauf.
\newblock Pattern search in flows based on similarity of stream line segments.
\newblock In {\em 19th International Workshop on Vision, Modeling and Visualization, VMV 2014, Darmstadt, Germany, 8-10 October 2014}, pp. 23--30, 2014.

\bibitem{Wang2016Patterns}
Z.~Wang, J.~M. Esturo, H.-P. Seidel, and T.~Weinkauf.
\newblock Stream line–based pattern search in flows.
\newblock {\em Computer Graphics Forum}, pp. n/a--n/a, 2016. \href{https://doi.org/10.1111/cgf.12990}
{doi: {{%
10\hspace{.1pt}\discretionary{.}{%
}{.}\hspace{.4pt}1111\discretionary{/}{%
}{/}cgf\hspace{.1pt}\discretionary{.}{%
}{.}\hspace{.4pt}12990}}}


\bibitem{wang2015multi}
Z.~Wang, H.-P. Seidel, and T.~Weinkauf.
\newblock Multi-field pattern matching based on sparse feature sampling.
\newblock {\em IEEE Transactions on Visualization and Computer Graphics}, 22(1):807--816, 2015.

\bibitem{wu2009topology}
K.~Wu, Z.~Liu, S.~Zhang, and R.~J. Moorhead~II.
\newblock Topology-aware evenly spaced streamline placement.
\newblock {\em IEEE Transactions on Visualization and Computer Graphics}, 16(5):791--801, 2009.

\bibitem{FlowInfo10}
L.~Xu, T.-Y. Lee, and H.-W. Shen.
\newblock An information-theoretic framework for flow visualization.
\newblock {\em IEEE Transactions on Visualization and Computer Graphics}, 16(6):1216--1224, 2010.

\bibitem{xu2010flow}
L.~Xu and H.-W. Shen.
\newblock Flow web: a graph based user interface for 3d flow field exploration.
\newblock In {\em Visualization and Data Analysis 2010}, vol. 7530, pp. 152--163. SPIE, 2010.

\bibitem{YuHierarchicalClustering}
H.~Yu, C.~Wang, C.-K. Shene, and J.~H. Chen.
\newblock Hierarchical streamline bundles.
\newblock {\em IEEE Transactions on Visualization and Computer Graphics}, 18(8):1353--1367, Aug. 2012. \href{https://doi.org/10.1109/TVCG.2011.155}
{doi: {{%
10\hspace{.1pt}\discretionary{.}{%
}{.}\hspace{.4pt}1109\discretionary{/}{%
}{/}TVCG\hspace{.1pt}\discretionary{.}{%
}{.}\hspace{.4pt}2011\hspace{.1pt}\discretionary{.}{%
}{.}\hspace{.4pt}155}}}


\bibitem{zafar2023extract}
A.~Zafar, D.~Yang, and G.~Chen.
\newblock Extract and characterize hairpin vortices in turbulent flows.
\newblock {\em IEEE Transactions on Visualization and Computer Graphics}, 2023.

\end{thebibliography}


\end{document}